\documentclass[11pt,oneside]{article}

\usepackage{epsfig,lscape}
\usepackage{amssymb,amsfonts,amsmath}
\usepackage{rotating}

\usepackage{amsthm}

\def\me{\mathrm e}

\def\dif{\mathrm d}

\def\var{\mathrm{var}}

\def\T{ {\mathrm{\scriptscriptstyle T}} }

\newtheorem{pro}{Proposition}

\theoremstyle{definition}

\theoremstyle{definition}

\begin{document}

\begin{titlepage}

\begin{center}
{\Large On doubly robust estimation for logistic partially linear models}

\vspace{.1in} Zhiqiang Tan\footnotemark[1]

\vspace{.1in}
\today
\end{center}

\footnotetext[1]{Department of Statistics, Rutgers University. Address: 110 Frelinghuysen Road,
Piscataway, NJ 08854. E-mail: ztan@stat.rutgers.edu. The research was supported in part by PCORI grant ME-1511-32740.}

\paragraph{Abstract.}
Consider a logistic partially linear model, in which the logit of the mean of a binary response is related to a linear function of some covariates
and a nonparametric function of other covariates. We derive simple, doubly robust estimators of coefficient for the covariates in the linear component of
the partially linear model.
Such estimators remain consistent if either a nuisance model is correctly specified for the nonparametric component,
or another nuisance model is correctly specified for the means of the covariates of interest given other covariates and the response at a fixed value.
In previous works, conditional density models are needed for the latter purposes unless a scalar, binary covariate is handled.
We also propose two specific  doubly robust estimators: one is locally-efficient like in our class of doubly robust estimators and
the other is numerically and statistically simpler and can achieve reasonable efficiency especially when the true coefficients are close to 0.

\paragraph{Key words and phrases.} Double robustness; Local efficiency; Logistic models; Odds ratio; Partially linear models; Semiparametric models.

\end{titlepage}

\section{Introduction} \label{sect:intro}

Generalized partially linear models are a semiparametric extension of generalized linear models (McCullagh \& Nelder 1989),
such that the conditional mean of a response variable $Y$ is related to a linear function of some covariates $Z$
and a smooth function of other covariates $X$. 
Let $\{(Y_i, Z_i, X_i): i=1,\ldots,n\}$ be independent and identically distributed observations from the joint distribution of $(Y,Z,X)$.
Consider the following model
\begin{align}
E ( Y | Z,X ) = \Psi \{ \beta^\T Z + g(X) \}, \label{eq:GPLM}
\end{align}
where $\Psi(\cdot)$ is an inverse link function, $\beta$ is a vector of unknown parameters, $g(\cdot)$ is an unknown, smooth function.
Estimation in such models has been studied in such models in at least two approaches.
In one approach, theory and methods have been developed in the case where $X$ is low-dimensional (for example, a scalar)
and kernel or spline smoothing is used to estimate $g(\cdot)$ at suitable rates of convergence (e.g., Speckman 1988; Severini \& Staniswalis 1994).
In another approach with $X$ relatively high-dimensional, doubly robust methods have been proposed to obtain estimators of $\beta$
which remain consistent and asymptotically normal at rate $n^{-1/2}$ if either a parametric model for $g(\cdot)$ or another parametric model about,
for example, $E(Z|X)$ is correctly specified (Robins \& Rotnitzky 2001; Tchetgen Tchetgen et al.~2010).

In this note, we are concerned with model (\ref{eq:GPLM}) with a binary response $Y$ (taking value 0 or 1) and a logistic link, hence a logistic partially linear model:
\begin{align}
P( Y=1 | Z,X ) = \mbox{expit} \{ \beta^\T Z + g(X) \}, \label{eq:logit-PLM}
\end{align}
where $\mbox{expit}(c) = \{1+\exp(-c)\}^{-1}$.
We provide a new class of doubly robust estimators of $\beta$ which remain consistent and asymptotically normal at rate $n^{-1/2}$ if
either a parametric model for $g(\cdot)$ or a parametric model for $E(Z| Y=0, X)$ is correctly specified,
under mild regularity conditions but without additional parametric or smoothness restriction.

Previously, doubly robust estimators of $\beta$ were derived in model (\ref{eq:GPLM}) with respect to parametric models for $g(\cdot)$ and $E (Z|X)$,
in the case of an identity  link, $\Psi(c) = c$, or a log link, $\Psi(c) = \exp(c)$  (Robins \& Rotnitzky 2001). For the logistic link, however,
no doubly robust estimator of $\beta$ can be constructed in this manner with respect to parametric models about $g(\cdot)$ and $E(Z|X)$ (Tchetgen Tchetgen et al.~2010).
In fact, doubly robust estimators of $\beta$ in model (\ref{eq:logit-PLM}) were obtained
with respect to parametric models about $g(\cdot)$ and $p(z|Y=0, X)$, the conditional density of $Z$ given $Y=0$ and $X$ (Chen 2007; Tchetgen Tchetgen et al.~2010).
Therefore, our result in general allows doubly robust estimation for $\beta$ in model (\ref{eq:logit-PLM})
with respect to more flexible nuisance models about the conditional mean $E(Z | Y=0, X)$ than about the conditional density $p(z| Y=0, X)$.
In the special case of binary $Z$, our class of doubly robust estimators of $\beta$ is equivalent to that in  Tchetgen Tchetgen et al.~(2010),
but involves use of the parametric model for $P(Z=1|Y=0,X)$ in a more direct manner.

We also propose two specific doubly robust estimators of $\beta$ in model (\ref{eq:logit-PLM}) based on efficiency considerations.
The first estimator requires numerical evaluation of expectations under a model for $p(z|Y=0, X)$
beyond the conditional mean $E( Z | Y=0, X)$ unless $Z$ is binary, but can be shown to
achieve the minimum asymptotic variance among our class of doubly robust estimators
when both models for $g(\cdot)$ and $p(z | Y=0, X)$ are correctly specified.
Compared with the locally efficient, doubly robust estimators in Tchetgen Tchetgen et al.~(2010),
this estimator remains consistent if the model for $p(z | Y=0, X)$ is misspecified but the less restrictive model for $E(Z | Y=0, X)$ is correctly specified.
Our second estimator is numerically and statistically simpler than our first one: it does not involve numerical integration or
a parametric specification of the conditional density $p(z|Y=0, X)$, and can achieve a similar asymptotic variance as our first estimator, especially when
the true value of $\beta$ is close to 0.

\section{Doubly robust estimation} \label{sec:DR}

For a semiparametric model, doubly robust estimation can often be derived by studying the orthogonal complement of the nuisance tangent space
(Robins \& Rotnitzky 2001).
Denote by $L_2$ the Hilbert space of $\dim(\beta)\times 1$ functions $q \equiv q(Y,Z,X)$, with the inner product defined as $E(q_1^\T q_2 )$.
Denote $\varepsilon^* = Y - \pi^*(Z,X)$, $\pi^*\equiv \pi^*(Z,X) = P(Y=1|Z,X)$, and
by $\beta^*$ and $g^* \equiv g^*(X)$ the truth of $\beta$ and $g(X)$.
For model (\ref{eq:logit-PLM}), the orthogonal complement of the nuisance tangent space is known to be
(Bickel et al.~1993; Robins \& Rotnitzky 2001)
\begin{align}
\Lambda^\bot
& = \left\{ \varepsilon^* \left(h - \frac{E[h\pi^*(1-\pi^*)|X]}{E[\pi^*(1-\pi^*)|X]} \right): h\equiv h(Z,X) \mbox{ unrestricted } \right\} \cap L_2. \label{eq:ortho-comp}
\end{align}
Our first result is a reformulation of $\Lambda^\bot$ as follows. See the Appendix for all proofs.

\begin{pro} \label{pro:ortho-comp}
Assume that $\pi^*(Z,X) \in (0,1)$ almost surely. The space $\Lambda^\bot$ can be equivalently expressed as
\begin{align}
\Lambda^\bot & = \left\{ \varepsilon^* \left(h - \frac{E[h\pi^* |Y=0, X]}{E[\pi^* |Y=0, X]} \right): h\equiv h(Z,X) \mbox{ unrestricted } \right\} \cap L_2\label{eq:ortho-comp2}\\
& = \Big\{ \zeta_0^* (u - E[u |Y=0, X]): u \equiv u(Z,X) \mbox{ unrestricted }  \Big\} \cap L_2 , \label{eq:ortho-comp3}
\end{align}
where $u \equiv u(Z,X)$ is a $\dim(\beta)\times 1$ function and
\begin{align*}
\zeta_0^*= \frac{\varepsilon^*}{\pi^*} = Y \frac{1-\pi^*}{\pi^*}-(1-Y) = Y \me^{ - \beta^{*\T}Z - g^*(X)} - (1-Y).
\end{align*}
\end{pro}

Our reformulation (\ref{eq:ortho-comp3}) suggests the following set of doubly robust estimating functions. Let $g(X;\alpha)$ be a parametric model for $g^*(X)$
and, independently, $f(X;\gamma)$ be a parametric model for $f^*(X) \equiv E (Z | Y=0, X)$.
The two functions $g^*(X)$ and $E(Z | Y=0, X)$ are variation independent, because $g^*(X)$ and $p(z|Y=0,X)$ are variation independent (Chen 2007).
For a $\dim(\beta) \times \dim(\beta)$ function $\phi(X)$, define
\begin{align}
r( Y, Z, X; \beta, \alpha, \gamma, \phi) = \left\{ Y \me^{ - \beta^\T Z - g(X; \alpha)} - (1-Y) \right\} \phi(X) \{ Z - f(X;\gamma) \}, \label{eq:DR-EE}
\end{align}
by letting $u(Z,X) = \phi(X) Z$ in (\ref{eq:ortho-comp3}).
Then $r( Y, Z, X; \beta, \alpha, \gamma, \phi)$ is an unbiased estimating function for $\beta^*$ if either model $g(X;\alpha)$ or $f(X;\gamma)$ is correctly specified.

\begin{pro} \label{pro:DR-EE}
If either $g^*(X)=g(X; \alpha)$ for some $\alpha$ or $f^*(X)=f(X;\gamma) $ for some $\gamma$, then
\begin{align*}
 E \{ r( Y, Z, X; \beta^*, \alpha, \gamma, \phi) \} = 0,
\end{align*}
provided that the above expectation exists.
\end{pro}

Various doubly robust estimators can be constructed through (\ref{eq:DR-EE}). In general, let $\hat\alpha$ be an estimator of $\alpha$,
for example, the maximum likelihood estimator, which satisfies  $\hat\alpha- \bar\alpha = n^{-1} \sum_{i=1}^n s_1 (Y_i, Z_i, X_i; \bar \alpha; \bar\beta) +o_p(n^{-1/2})$
for some constant $(\bar \alpha, \bar\beta)$ and influence function $s_1(\cdot)$ such that
$g(X; \bar \alpha) = g^*(X)$ if model $g(X;\alpha)$ is correctly specified. 
Let $\hat\gamma$ be an estimator of $\gamma$,
for example, the least-squares or related estimator, which satisfies  $\hat\gamma- \bar\gamma = n^{-1} \sum_{i=1}^n s_2 (Y_i, Z_i, X_i; \bar \gamma) $ $ +o_p(n^{-1/2})$
for some constant $\bar \gamma$ and influence function $s_2(\cdot)$ such that
$f(X; \bar \gamma) = f^*(X)$ if model $f(X;\gamma)$ is correctly specified. 
Define an estimator $\hat\beta (\phi)$ as a solution to
\begin{align*}
\frac{1}{n} \sum_{i=1}^n r( Y_i, Z_i, X_i; \beta, \hat\alpha, \hat\gamma, \phi)= 0 .
\end{align*}
Under suitable regularity conditions (e.g., Manski 1988), it can be shown that if either model $g(X;\alpha)$ or $f(x;\gamma)$ is correctly specified, then
\begin{align}
\hat \beta(\phi) - \beta^* & = \frac{H^{-1}}{n} \sum_{i=1}^n  \Big\{ r( Y_i, Z_i, X_i; \beta^*, \bar\alpha, \bar\gamma, \phi) \nonumber \\
& \quad - B_1 s_1(Y_i,Z_i,X_i; \bar\alpha, \bar\beta) -B_2 s_2(Y_i, Z_i, X_i; \bar\gamma) \Big\} + o_p(n^{-1/2}), \label{eq:DR-IF}
\end{align}
where $H = E\{ \partial r(Y,Z,X; \beta,\bar\alpha,\bar\gamma,\phi)/\partial \beta\} |_{\beta=\beta^*}$,
$B_1 = E \{ \partial r(Y,Z,X; \beta^*,\alpha,\bar\gamma,\phi) /\partial\alpha\} |_{\alpha=\bar\alpha}$,\
and $B_1 = E \{ \partial r(Y,Z,X; \beta^*,\bar\alpha, \gamma,\phi) /\partial\gamma\} |_{\gamma=\bar\gamma}$.
The asymptotic variance of $\hat\beta (\phi)$ can be estimated by using the sample variance of an estimated version of the influence function in (\ref{eq:DR-IF}).

We now provide several remarks. First, estimating function (\ref{eq:DR-EE}) can be expressed as
\begin{align}
r( Y, Z, X; \beta, \alpha, \gamma, \phi) = \left\{ \frac{Y}{\pi(Z,X ; \beta,\alpha)} -1 \right\} \phi(X) \{ Z - f(X;\gamma) \}, \label{eq:DR-EE2}
\end{align}
where $\pi(Z,X ; \beta,\alpha) = \mbox{expit} \{\beta^\T Z + g(X;\alpha)\}$, representing the conditional probability $P(Y=1|Z,X)$ under the conjunction of
model (\ref{eq:logit-PLM}) and model $g(X;\alpha)$.
Therefore, our doubly robust estimating function involves the product of two ``residuals", $\pi^{-1}(Z,X ; \beta,\alpha) Y -1$ and $Z - f(X;\gamma)$.
Similar products can also be found in previous doubly robust estimating functions for $\beta$ in model (\ref{eq:GPLM})
with the identity or log link (Robins \& Rotnitzky 2001). However,
a notable feature in (\ref{eq:DR-EE2}) is that
the residual used from the model $P(Y=1 | Z,X) = \pi(Z,X; \beta,\alpha)$ is $\pi^{-1}(Z,X ; \beta,\alpha) Y -1$, associated with the estimating equation for
calibrated estimation (Tan 2017), which in the case $g(X;\alpha) = \alpha^\T X$ gives
\begin{align*}
\frac{1}{n} \sum_{i=1}^n \left\{ \frac{Y}{\pi(Z,X ; \beta,\alpha)}-1 \right\} (Z^\T,X^\T)^\T =0 .
\end{align*}
The standard residual from logistic regression is $Y - \pi(Z,X ; \beta,\alpha)$, associated with the score equation
for maximum likelihood estimation, which in the case $g(X;\alpha) = \alpha^\T X$ gives
\begin{align*}
\frac{1}{n} \sum_{i=1}^n \{ Y - \pi(Z,X ; \beta,\alpha) \} (Z^\T,X^\T)^\T =0 .
\end{align*}
In general, the estimating function $\{Y - \pi(Z,X ; \beta,\alpha) \} \phi(X) \{ Z - f(X;\gamma) \}$ is  not unbiased for $\beta^*$ if model $f(X;\gamma)$ is correctly specified
but model $g(X;\alpha)$ is misspecified.

Second, our results can also be used to shed light on the class of doubly robust estimators in Tchetgen Tchetgen et al.~(2010), which are briefly reviewed as follows.
For model (\ref{eq:logit-PLM}), the conditional distribution of $(Y,Z)$ jointly given $X$ can be determined as (Chen 2007)
\begin{align}
p(y,z |X ) = c^{-1}(X) \me^{\beta^\T (z-z_0) y} p(z | Y=0,X) p(y | Z=z_0, X), \label{eq:joint-dens}
\end{align}
where $z_0$ is some fixed value (assumed to be 0 hereafter), $c(X) = \int  \me^{\beta^\T zy} p(z | Y=0,X) p(y | Z=z_0, X) \,\dif \mu(z,y)$, and
the conditional densities $p(z| Y=0,X)$ and $p(y |Z=0, X)$ are variation-independent nuisance parameters.
Let $p^\dag(y,z|X) = p_1^\dag(y|X) p_2^\dag (z|X)$ be some pre-specified conditional densities $p_1^\dag (y|X)$ and $p_2^\dag (z|X)$.
By using (\ref{eq:joint-dens}), the ortho-complement of the nuisance tangent space in model (\ref{eq:logit-PLM}) can be characterized as (Tchetgen Tchetgen et al.~2010)
\begin{align}
\Lambda^\bot = \left\{ [d(Y,Z,X) - d^\dag(Y,Z,X)] \frac{p^\dag(Y,Z|X)}{p(Y,Z|X)}:  d(Y,Z,X) \mbox{ unrestricted } \right\} \cap L_2, \label{eq:ortho-comp4}
\end{align}
where $d^\dag (Y,Z,X) = E^\dag ( D |Z,X) - E^\dag (D |Y,X) - E^\dag ( D| X)$ for $D \equiv d(Y,Z,X)$, and
$E^\dag ( \cdot | \cdot, X)$ denotes the expectation under $ p^\dag(y,z|X)$.
It can be verified by direct calculation that the two sets on the right hand sides of (\ref{eq:ortho-comp}) and (\ref{eq:ortho-comp4}) are equivalent to each other:
each element in the right hand side of (\ref{eq:ortho-comp4}) can be expressed in the form of elements in the right hand side of (\ref{eq:ortho-comp}),
and vice versa.
Let $p(y | Z=0, X; \alpha)$ or equivalently $g(X;\alpha)$ be a parametric model for $p(y | Z=0, X)$ or $g^*(X)$, and let
$p(z | Y=0, X; \theta)$ be a parametric model for $p(z |Y=0,X)$.
For a $\dim(\beta)\times 1$ function $h \equiv h(Z,X)$, the estimating function based on (\ref{eq:ortho-comp4}) in Tchetgen Tchetgen et al.~(2010) can be equivalently defined, based on (\ref{eq:ortho-comp}), as
\begin{align}
\tau ( Y,Z,X ; \beta,\alpha, \theta, h) = \{ Y - \pi(Z,X; \beta,\alpha) \} \left\{ h(Z,X) - \frac{ E[h \pi(1-\pi)| X ;\beta,\alpha,\theta]}{ E[ \pi(1-\pi)|X; \beta,\alpha,\theta]} \right\}, \label{eq:DR-TRR}
\end{align}
where $\pi \equiv \pi(Z,X; \beta,\alpha)= \mbox{expit} \{\beta^\T Z + g(X;\alpha)\}$ and $E(\cdot| X; \beta,\alpha,\theta)$ denotes the expectation under
the law defined as (\ref{eq:joint-dens}), but evaluated at $p(y | Z=0, X; \alpha)$ and $p(z | Y=0, X; \theta)$.
The estimating function (\ref{eq:DR-TRR}) is doubly robust, i.e.~unbiased for $\beta^*$ if either model
$p(y | Z=0, X; \alpha)$  or $p(z | Y=0, X; \theta)$ is correctly specified.
Although (\ref{eq:DR-TRR}) appears to be asymmetric in $Y$ and $Z$, the double robustness of (\ref{eq:DR-TRR}) follows from that of its equivalent version based on (\ref{eq:ortho-comp4}),
as shown by exploiting the symmetry in $Y$ and $Z$ in Tchetgen Tchetgen et al.~(2010).
See also Tchetgen Tchetgen and Rotnitzky (2011) for an explicit demonstration of symmetry of (\ref{eq:DR-TRR}) in $Y$ and $Z$ with $h(Z,X)=Z$
in the case of a binary $Z$.

As an interesting implication of our reformulation (\ref{eq:ortho-comp2}) in Proposition~\ref{pro:ortho-comp}, the estimating function (\ref{eq:DR-TRR}) can be equivalently expressed as
\begin{align}
\tau ( Y,Z,X ; \beta,\alpha, \theta, h) = \{ Y - \pi(Z,X; \beta,\alpha) \} \left\{ h(Z,X) - \frac{ E[h \pi | Y=0, X ; \theta]}{ E[ \pi |Y=0,X; \theta]} \right\}, \label{eq:DR-TRR2}
\end{align}
which involves the expectation $E(\cdot | Y=0, X; \alpha)$ under $p(z | Y=0, X; \theta)$, instead of $E(\cdot| X; \beta,\alpha,\theta)$ under
the law (\ref{eq:joint-dens}) evaluated at $p(y | Z=0, X; \alpha)$ and $p(z | Y=0, X; \theta)$.
Therefore, (\ref{eq:DR-TRR2}) is computationally much simpler than (\ref{eq:DR-TRR}) and its equivalent version based on (\ref{eq:ortho-comp4}).
Moreover, the double robustness of (\ref{eq:DR-TRR2})  with respect to $p(y | Z=0, X; \alpha)$  and $p(z | Y=0, X; \theta)$ can be directly shown as in the Appendix, without
invoking its equivalent version based on (\ref{eq:ortho-comp4}).

Third, we compare our doubly robust estimating functions with those in Tchetgen Tchetgen et al.~(2010).
For a $\dim(\beta)\times 1$ function $u \equiv u(Z,X)$, consider the estimating funtion
\begin{align}
\tau^\prime ( Y,Z,X ; \beta,\alpha, \theta, u) = \left\{ \frac{Y}{\pi(Z,X; \beta,\alpha)} -1 \right\} \left\{ u(Z,X) - E[u  | Y=0,X ;\theta]  \right\}. \label{eq:DR-TRR3}
\end{align}
By our reformulation (\ref{eq:ortho-comp3}), the class of estimating functions $\tau^\prime ( Y,Z,X ; \beta,\alpha, \theta, h) $
over all possible choices of $u(Z,X)$ is equivalent to that of $\tau ( Y,Z,X ; \beta,\alpha, \theta, h)$ over all possible choices of $h(Z,X)$ as used in Tchetgen Tchetgen et al.~(2010).
A subtle point is that the mapping between $h(Z,X)$ and $u(Z,X)$ depends on $\pi(Z, X; \beta,\alpha)$, but this does not affect our subsequent discussion.
Similarly as (\ref{eq:DR-TRR2}), the estimating function (\ref{eq:DR-TRR3}) can be shown to be doubly robust for $\beta^*$ with respect to models
$p(y | Z=0, X; \alpha)$ and $p(z | Y=0, X; \theta)$.

By comparing (\ref{eq:DR-EE}) and (\ref{eq:DR-TRR3}), we see that our estimating function (\ref{eq:DR-EE}) corresponds to a particular choice of estimating function (\ref{eq:DR-TRR3})
with $u(Z,X) = \phi(X) Z$, such that (\ref{eq:DR-EE}) depends only on a parametric model for the conditional expectation $E( Z | Y=0, X)$, but not the conditional density $p(z| Y=0,X)$.
Therefore, our class of (\ref{eq:DR-EE}) is in general a strict subset of the class of (\ref{eq:DR-TRR3}) to achieve double robustness with respect to conditional mean models for $E(Z |Y=0,X)$,
except when $Z$ is binary and hence the classes of (\ref{eq:DR-EE}) and (\ref{eq:DR-TRR3}) are equivalent.

Fourth, there is a similar characterization of $\Lambda^\bot$ as in Proposition~\ref{pro:ortho-comp}, involving expectations under $p(z|Y=1,X)$ instead of $p(z|Y=0,X)$.
By symmetry, it can be shown that
\begin{align*}
\Lambda^\bot & = \left\{ \varepsilon^* \left(h - \frac{E[h(1-\pi^*) |Y=1, X]}{E[1-\pi^* |Y=1, X]} \right): h\equiv h(Z,X) \mbox{ unrestricted } \right\} \cap L_2 \\
& = \Big\{ \zeta_1^* (u - E[u |Y=1, X]): u \equiv u(Z,X) \mbox{ unrestricted }  \Big\} \cap L_2 ,
\end{align*}
where $u \equiv u(Z,X)$ is a $\dim(\beta)\times 1$ function and $\zeta_1^*= \varepsilon^* /(1-\pi^*) = Y - (1-Y)\me^{ \beta^{*\T}Z + g^*(X)}$.
Consequently, a similar estimating function as (\ref{eq:DR-EE}) can be derived such that it is doubly robust for $\beta^*$
with respect to parametric models for $g^*(X)$ and $E(Z| Y=1, X)$.

\section{Efficiency considerations}

For our class of doubly robust estimating functions (\ref{eq:DR-EE}), we study how to choose the function $\phi(X)$ based on efficiency considerations.
First, the following result gives the optimal choice of $\phi(X)$ with correctly specified models $g(X;\alpha)$ and $f(X;\gamma)$.

\begin{pro} \label{pro:efficiency}
If both models $g(X;\alpha)$ and $f(X;\gamma)$ are correctly specified for $g^*(X)$ and $E(T|Y=0,X)$ respectively, then
the optimal choice of $\phi(X)$ in minimizing the asymptotic variance of $\hat\beta (\phi)$ which admits asymptotic expansion (\ref{eq:DR-IF}) is
\begin{align*}
\phi_{\mbox{\tiny opt}}(X) & =  E [ (Z - E(Z|Y=0,X))^{\otimes 2} |Y=0,X ] \\
& \quad \times E^{-1}[ \pi^{*-1}(Z,X) (Z - E(Z|Y=0,X))^{\otimes 2} | Y=0, X ] ,
\end{align*}
where $b^{\otimes 2} = b b^\T$ for a column vector $b$.
\end{pro}

From this result, it is straightforward to derive a locally-efficient like, doubly robust estimator for $\beta^*$.
Let $(\hat\beta, \hat\alpha)$ be the maximum likelihood estimator in the model $\pi(Z,X ; \beta,\alpha) = \mbox{expit} \{\beta^\T Z + g(X;\alpha)\}$,
and $\hat\theta$ be the maximum likelihood estimator in a conditional density model $p(z | Y=0, X; \theta)$ as in (\ref{eq:DR-TRR}) but
compatible with model $f(X;\gamma)$ for $E(Z | Y=0, X)$, where $\theta= (\gamma, \gamma^\prime)$ and $\gamma^\prime$ is a variance parameter.
Consider the estimator $\hat\beta (\hat \phi_{\mbox{\tiny opt}})$ with
\begin{align*}
\hat \phi_{\mbox{\tiny opt}}(X) & = E [ ( Z - f(X;\hat\gamma)^{\otimes 2} ) |Y=0,X; \hat\theta ]\\
& \quad \times  E^{-1}[ \pi^{-1}(Z,X; \hat\beta,\hat\alpha) (Z - f(X;\hat\gamma))^{\otimes 2} | Y=0, X; \hat\theta ]  .
\end{align*}
Then it can be shown under suitable regularity conditions that
$\hat\beta (\hat \phi_{\mbox{\tiny opt}})$ is doubly robust, i.e.~remains consistent for $\beta^*$ if either model $g(X;\alpha)$ or $f(X;\gamma)$ is correctly specified,
and achieves the minimum asymptotic variance among all estimators $\hat\beta(\phi)$ when both models $g(X;\alpha)$ and $p(z | Y=0, X; \theta)$ including $f(X; \gamma)$ are correctly specified.

It is interesting to compare  $\hat\beta (\hat \phi_{\mbox{\tiny opt}})$ with the locally efficient, doubly robust estimator for $\beta^*$ in Tchetgen Tchetgen et al.~(2010).
For a $\dim(\beta)\times 1$ function $h(Z,X)$, define an estimator $\hat\beta(h)$ as a solution to
$n^{-1} \sum_{i=1}^n \tau(Y,Z,X ; \beta, \hat\alpha, \hat\theta,  h)=0$,
where $(\hat\alpha, \hat \theta)$ are maximum likelihood estimators as above or, without affecting our discussion here, profile maximum likelihood estimators as in Tchetgen Tchetgen et al.~(2010).
Then the optimal choice of $h(Z,X)$ in minimizing the asymptotic variance of $\hat\beta (h)$
is $h_{\mbox{\tiny eff}}(Z,X) = \partial (\beta^\T Z) /\partial \beta = Z$.
In fact, the estimator $\hat\beta (h_{\mbox{\tiny eff}})$ is locally efficient, i.e.~achieving the semiparametruc variance bound in model (\ref{eq:logit-PLM})
when both models $g(X;\alpha)$ and $p(z | Y=0, X; \theta)$ are correctly specified.
Unless $Z$ is binary, this semiparametric variance bound is in general strictly smaller than the asymptotic variance achieved by $\hat\beta (\hat \phi_{\mbox{\tiny opt}})$ when both
models $g(X;\alpha)$ and $p(z | Y=0, X; \theta)$ are correctly specified,
because the class of estimating functions (\ref{eq:DR-EE}) is strictly a subset of the class (\ref{eq:DR-TRR}), (\ref{eq:DR-TRR2}), or (\ref{eq:DR-TRR3}),
as discussed in Section~\ref{sec:DR}.
In the case of a binary $Z$ and hence $\theta=\gamma$, the two estimators $\hat\beta (\hat \phi_{\mbox{\tiny opt}})$ and $\hat\beta (h_{\mbox{\tiny eff}})$ are equivalent.
On the other hand, $\hat\beta (h_{\mbox{\tiny eff}})$ is doubly robust only with respect to models $g(X;\alpha)$ and $p(z| Y=0, X; \theta)$,
whereas $\hat\beta (\hat \phi_{\mbox{\tiny opt}})$ is doubly robust with respect to $g(X;\alpha)$ and $f(X;\gamma)$ and hence
remains consistent for $\beta^*$ if model $p(z| Y=0, X; \theta)$ is misspecified but
the less restrictive model $f(X;\gamma)$ for $E(Z|Y=0,X)$ is correctly specified.

Evaluation of the function $\hat \phi_{\mbox{\tiny opt}}(X)$ and hence the estimator $\hat\beta (\hat \phi_{\mbox{\tiny opt}})$ in general requires cumbersome numerical integration
with respect to the density $p(z | Y=0, X; \hat\theta)$.
For computational simplicity, consider the estimator $\hat\beta (\phi_{\mbox{\tiny simp}})$ with scalar
$\phi_{\mbox{\tiny simp}} (X) = P (Y=1 | Z=0, X; \hat \alpha) = \mbox{expit} \{ g(X; \hat \alpha) \}$.
The corresponding estimating function can be shown to become
\begin{align}
r ( Y,Z, X; \beta, \hat \alpha, \hat \gamma, \phi_{\mbox{\tiny simp}}) =
\frac{ Y \me^{-\beta^\T Z} - (1-Y) \me^{g(X; \hat\alpha)} }{1 +\me^{g(X; \hat\alpha)}} \{ Z - f(X; \hat\gamma)\}. \label{eq:DR-EE-simp}
\end{align}
The particular choice $\phi_{\mbox{\tiny simp}}(X)$ can be motivated by the fact that
if the true $\beta^*=0$ then $\phi_{\mbox{\tiny opt}}(X) =  \mbox{expit} \{ g^*(X) \}$.
Then $\hat\beta (\phi_{\mbox{\tiny simp}})$ is nearly as efficient as $\hat\beta (\hat \phi_{\mbox{\tiny opt}})$
and, by similar reasoning, also $\hat\beta (h_{\mbox{\tiny eff}})$
whenever $\beta^*$ is close to 0. This is analogous to how the easy-to-compute estimator is related to the locally efficient estimator $\hat\beta(h_{\mbox{\tiny eff}})$
in Tchetgen Tchetgen et al.~(2010, Section 4).
Moreover, the estimating function (\ref{eq:DR-EE-simp}) can be equivalently expressed as
\begin{align*}
r ( Y,Z, X; \beta, \hat \alpha, \hat \gamma, \phi_{\mbox{\tiny simp}}) =
\me^{-\beta^\T Z Y} [ Y-\mbox{expit} \{ g(X; \hat\alpha) \} ]  \{ Z - f(X; \hat\gamma)\} , \label{eq:DR-EE-simp}
\end{align*}
which, in the case of a binary $Z$, coincides with the estimating function underlying the closed-form estimator for $\beta^*$ in Tchetgen Tchetgen (2013).

\section{Conclusion}

We derive simple, doubly robust estimators of coefficients for the covariates in the linear component in a logistic partially linear model.
Such estimators remain consistent if either a nuisance model is correctly specified for the nonparametric component of the partially linear model,
or a conditional mean model is correctly specified for the covariates of interest given other covariates and the response at a fixed value.
These estimators can be useful in conventional settings with a limited number of covariates. Moreover, there have been various works exploiting
doubly robust estimating functions to obtain valid inferences in high-dimensional problems (e.g., Farrell 2015; Chernozhukov et al.~2018; Tan 2018).
Our estimating functions can potentially be employed to achieve similar properties in high-dimensional settings.

\section{Appendix}

{\noindent \textbf{Proof of Proposition~\ref{pro:ortho-comp}.}} {
First, we show that for any $h \equiv h(Z,X)$,
\begin{align*}
E [ h \pi^* (1-\pi^*) |X ] = P(Y=0|X)  E [ h \pi^* | Y=0, X] .
\end{align*}
This follows because
$ E [ h \pi^* (1-\pi^*) |X ]  = E [ h \pi^* 1\{Y=0\} | X ] = P(Y=0|X)  E [ h \pi^* | Y=0, X]$
by the law of iterated expectations and then the law of total probability.
Then the set (\ref{eq:ortho-comp}) is equivalent to (\ref{eq:ortho-comp2}).
Next, the set (\ref{eq:ortho-comp2}) is equivalent to $\{ \varepsilon^* h_c : h_c \equiv h_c(Z,X) \mbox{ satisfying } E [h_c \pi^* | Y=0, X] = 0 \} \cap L_2$,
and the set (\ref{eq:ortho-comp3}) is equivalent to
$\{ \zeta_0^* u_c : u_c \equiv u_c(Z,X) \mbox{ satisfying } E [ u_c | Y=0, X] = 0 \} \cap L_2 $.
The two sets are equivalent to each other, by letting $h_c = u_c \pi^*$.
\hfill $\Box$ \vspace{.1in}}

{\noindent \textbf{Proof of Proposition~\ref{pro:DR-EE}.}} {
By the law of iterated expectations, we have
\begin{align*}
& E \{ r( Y, Z, X; \beta^*, \alpha, \gamma, \phi) \}
= E \left[ E \left\{ Y \me^{ - \beta^{*\T} Z - g(X;\alpha)} - (1-Y) \bigg| Z,X\right\} \phi(X) \{ Z - f(X;\gamma)\} \right] \\
& \quad = E \left[ (1-Y) \left\{ \me^{ g^*(X) - g(X;\alpha)} - 1 \right\} \phi(X) \{ Z - f(X;\gamma)\} \right] .
\end{align*}
This immediately shows that if either $g(X; \alpha) = g^*(X)$ or $f(X;\gamma) = f^*(X)$, then $E \{ r( Y, Z, X; $ $\beta^*, \alpha, \gamma, \phi) \}=0$.
\hfill $\Box$ \vspace{.1in}}

{\noindent \textbf{Proof of double robustness of (\ref{eq:DR-TRR2}).}} {
By the law of iterated expectations, we have
\begin{align*}
& E \{ \tau( Y, Z, X; \beta^*, \alpha, \theta, h) \}
= E \left[ E \left\{ Y \me^{ - \beta^{*\T} Z - g(X;\alpha)} - (1-Y) \bigg| Z,X\right\} \left\{ h\pi - \frac{\pi E[h \pi | Y=0, X ; \theta]}{ E[ \pi |Y=0,X; \theta]} \right\}\right] \\
& \quad = E \left[ (1-Y) \left\{ \me^{ g^*(X) - g(X;\alpha)} - 1 \right\} \left\{ h\pi - \frac{\pi E[h \pi | Y=0, X ; \theta]}{ E[ \pi |Y=0,X; \theta]} \right\} \right] .
\end{align*}
This immediately shows that if either $g(X; \alpha) = g^*(X)$ or $p(z|Y=0,X;\theta) = p(z|Y=0,X)$, then $E \{ \tau( Y, Z, X; $ $\beta^*, \alpha, \theta, \phi) \}=0$.
\hfill $\Box$ \vspace{.1in}}

{\noindent \textbf{Proof of Proposition~\ref{pro:efficiency}.}} {
Suppose that both models $g(X;\alpha)$ and $f(X;\gamma)$ are correctly specified, such that $g(X;\bar\alpha) = g^*(X)$ and $f(X;\bar\gamma) = E(T|Y=0,X)$. Then
$B_1=B_2=0$ by direct calculation, and hence (\ref{eq:DR-IF}) reduces to
\begin{align*}
\hat \beta(\phi) - \beta^* & = \frac{H^{-1}}{n} \sum_{i=1}^n  r( Y_i, Z_i, X_i; \beta^*, \bar\alpha, \bar\gamma, \phi)  + o_p(n^{-1/2}) .
\end{align*}
By the proof of Proposition~\ref{pro:DR-EE}, we actually have $E \{\varrho(Y,Z,X; \beta^*)|X\}=0$,
where \begin{align*}
\varrho(Y,Z,X; \beta) = \left\{ Y \me^{ - \beta^\T Z - g(X; \bar\alpha)} - (1-Y)\right\}  \{ Z - f(X;\bar \gamma)\}.
\end{align*}
Therefore, $\hat\beta(\phi)$ is asymptotically equivalent to a solution to $n^{-1} \sum_{i=1}^n \phi(X_i) \varrho( Y_i, Z_i, X_i; \beta)=0$,
which can be seen as an estimator for $\beta^*$ under the conditional moment condition $E \{\varrho(Y,Z,X;$ $ \beta^*)|X\}=0$.
By Chamberlain (1987), the optimal choice of $\phi(X)$ in minimizing the asymptotic variance of such an estimator is
$E^\T\{ \partial \rho(Y,Z,X ; \beta) /\partial \beta^\T |X \} |_{\beta=\beta^*} \var^{-1} \{ \rho(Y,Z,X; \beta^*) |X \}$,
which can be simplified as $\phi_{\mbox{\tiny opt}}(X)$ by direct calculation.
\hfill $\Box$ \vspace{.1in}}

\vspace{.15in}
\centerline{\bf\Large References}

\begin{description}\addtolength{\itemsep}{-.1in}

\item Bickel, P.J., Klaassen, C.A.J., Ritov, Y., and Wellner, J.A. (1993) {\em Efficient
and Adaptive Estimation for Semiparametric Models}, The Johns Hopkins University Press, Baltimore.

\item Chamberlain, G. (1987) ``Asymptotic efficiency in estimation with conditional
moment restrictions," {\em Journal of Econometrics}, 34, 305-334.

\item Chen, H.Y. (2007) ``A semiparametric odds ratio model for measuring association, {\em Biometrics}, 63, 413-421.

\item Chernozhukov, V., Chetverikov, D., Demirer, M., Duflo, E., Hansen, C., Newey, W.K., and Robins, J.M. (2018)
``Double/debiased machine learning for treatment and structural parameters," {\em Econometrics Journal}, 21, C1-C68.

\item Farrell, M.H. (2015) ``Robust inference on average treatment effects with possibly more covariates
than observations." {\em Journal of Econometrics}, 189, 1--23.


\item Manski, C.F. (1988) {\em Analog Estimation Methods in Econometrics}, Chapman \& Hall, New York

\item McCullagh, P. and Nelder, J.A. (1989) {\em Generalized Linear Models} (2nd edition), Chapman \& Hall, London.

\item Robins, J.M., and Rotnitzky, A. (2001) Comment on the Bickel and Kwon Article, ``Inference for semiparametric models: Some questions and an
answer," {\em Statistica Sinica}, 11, 920-936.

\item Severini, T.A. and Staniswalis, J.G. (1994) ``Quasi-likelihood estimation in semiparametric models," {\em
Journal of the American Statistical Association}, 89, 501-511.

\item Speckman, P. (1988) ``Kernel smoothing in partial linear models," {\em Journal of the Royal Statistical Society}, Ser.~B, 50, 413-436.

\item Tan, Z. (2017) ``Regularized calibrated estimation of propensity scores with model misspecification and high-dimensional data," arXiv:1710.08074.

\item Tan, Z. (2018) ``Model-assisted inference for treatment effects using regularized calibrated estimation with high-dimensional data," arXiv:1801.09817.

\item Tchetgen Tchetgen, E.J. (2013) ``On a closed-form doubly robust estimator of the adjusted odds ratio for a
binary exposure," {\em American Journal of Epidemiology}, 177, 1314-1316.

\item Tchetgen Tchetgen E.J. and Rotnitzky A. (2011) ``Double-robust estimation
of an exposure-outcome odds ratio adjusting for confounding in
cohort and case-control studies," {\em Statistics in Medicline}, 30, 335-347.

\item Tchetgen Tchetgen, E.J., Robins, J.M., and Rotnitzky, A. (2010) ``On doubly robust estimation in a semiparametric odds ratio model,”
{\em Biometrika}, 97, 171-180.
\end{description}

\end{document}